\documentclass[]{spie}  
\usepackage[]{graphicx}

\usepackage[]{amsmath}

\title{Precipitable Water Vapor: Considerations on the water vapor scale height, dry bias of the radiosonde humidity sensors, and spatial and temporal variability of the humidity field}

\author{Angel C. Otarola\supit{a}, Richard Querel\supit{b}, and Florian Kerber\supit{c}
\skiplinehalf
\supit{a}Thirty Meter Telescope, 1111 South Arroyo Parkway, Suite 200, Pasadena, California, USA \\
\supit{b} Dpto. Ingenier{\'i}a El{\'e}ctrica, Universidad De Chile, Av. Tupper 2007, Santiago, Chile\\
\supit{c} European Southern Observatory, Karl-Schwarzschild-Str. 2, 85748 Garching, Germany \\
}

\authorinfo{Further author information: (Send correspondence to Angel C. Otarola): E-mail: aotarola@tmt.org, Telephone: 1 520 318 8401}

\begin{document}
  \maketitle

\begin{abstract}

The Thirty Meter Telescope (TMT) and the European Extremely Large Telescope (E-ELT) site testing teams have recently finalized their site testing campaigns to determine the best site for their respective deployments. Since atmospheric water vapor is the dominant source of absorption and increased thermal background in the infrared, and in particular in the mid-infrared, both projects included precipitable water vapor (PWV) measurements in their corresponding site testing campaigns. TMT originally planned to monitor PWV at the sites of interest by means of using infrared radiometers operating at 20 $\mu$m.  Technical failures and calibration issues prevented them from having a sufficiently long PWV time-series to properly characterize the sites using this method.  Therefore, they used surface measurements of temperature and relative humidity to derive surface water vapor density, that, taken together with an assumed water vapor scale height, allowed for the estimation of integrated water vapor in the atmospheric column. In this way, surface data were used to understand the long-term median value of PWV at the TMT candidate sites in northern Chile. On the other hand, the E-ELT team conducted dedicated PWV measurement campaigns at two of their observatory sites using radiosonde soundings to validate historical time-series of PWV reconstructed by way of a spectroscopic analysis of astronomical standard sources observed with the FEROS instrument at the La Silla site, and with UVES, CRIRES and VISIR at the Paranal site.  The E-ELT also estimated the median PWV for the Armazones site from extrapolation of their Paranal statistics and accounting for the difference in elevation between the two sites; as well as an archival analysis of radiosonde data available from the city of Antofagasta by integration of the humidity profile starting from 3000 m altitude. In the case of the Armazones site, the published median PWV by both groups differ by about 1 mm (approximately 30\% difference), with the E-ELT values being drier than those estimated by the TMT group. This work looks at some of the possible reasons that could explain this difference, among them the water vapor scale height (needed in the case of the TMT study), the horizontal variability of the water vapor field, and an unaccounted correction due to a dry bias known to affect the radiosondes relative humidity sensors in the case of the E-ELT study.
\end{abstract}



\section{INTRODUCTION}
\label{sec:intro}  

The existing large aperture telescope projects have important science drivers for the near- and mid-infrared  (NIR, MIR) spectral bands.  The interesting scientific drivers where MIR is particularly helpful are, for instance: the study of star formation and the evolution of circumstellar disks, detection of organic molecules (eg. in planet forming regions), the physical conditions and structures in Active Galactic Nuclei (AGN), the study of highly redshifted Gamma Ray Bursts (GRB), and the detection of direct thermal emission from exo-planets, and several others\cite{eelt_science_cases}. Water vapor in the Earth's atmosphere is the dominant source of absorption of radiation and increased thermal background in the MIR spectral band\cite{chamberlain2000}, and consequently affects the performance of astronomy research from all ground-based astronomical facilities.

The geographic zones of interest for astronomy are those that are both dry and offer a large number of cloud-free nights\cite{graham2004}. With the exemption of the polar regions, where the dryness of the atmosphere results from the extremely low temperatures\cite{saunders2009,eric2010,curry1995} and the Clausius-Clapeyron relationship; several other regions, dominated by high pressure centers with the subduction of relatively dry air, have been found to meet the conditions for the deployment of astronomical facilities (see Ref. 3 and 7). In these geographic areas water vapor typically decreases with altitude, and consequently the higher the location the drier the air (see Figure 8 in Ref. 8), but also the colder due to adiabatic expansion of the troposphere that gives rise to a temperature lapse rate of about -6.5 K/km. This implies less absorption and lower atmospheric thermal background emission, thus improving the performance of astronomical imaging and spectroscopy in the MIR band. However, with respect to atmospheric water vapour, the Earth's atmosphere is highly unsaturated, which gives rise to the high degree of both spatial and temporal variability observed in its distribution.  For these reasons, the overall distribution, variability and seasonality of water vapor above astronomical sites of interest are currently being studied.

The principal methods of measuring precipitable water vapor in the atmospheric column at sites of interest are by means of: satellite platform sensors such as GOES Imager 6.7 $\mu$m channel\cite{erasmus2000} or from reflected solar radiation in selected bands in the wavelength\footnote{Typically in the visible through far infrared the authors prefer to speak in terms of wavelength, while in the sub-millimeter through microwave spectral bands the spectral characteristics are mainly given in terms of frequency of the radiation.} range from 390 nm to 1040 nm as is the case of the ENVISAT MERIS imaging spectrometer\cite{kurlan2007}, from absorption of weak telluric water lines\cite{thomas_osip2007},  from the change of the continuum level with atmospheric water concentration of standard stars observed with astronomical spectrographs\cite{smette2007,kerber2010a,querel2011}, ground-based radiometry at spectral bands sensitive to atmospheric water vapor absorption such as 220 GHz, 225 GHz\cite{kono1995,mckinnon1987,radford2002}, 210 GHz\cite{hiriart2003,otarola2009} and 215 GHz\cite{hiriart1997}, 183 GHz\cite{delgado1999}, 20 $\mu$m\cite{smith2001}, by integration of the humidity profile obtained by radiosonde soundings\cite{giovanelli2001,chacon2010}, and also from surface measurements of water vapor density together with the assumption of an exponential decrease of water vapor with a given scale height\cite{otarola2010a}. Another technique is that of estimating PWV from the wet-path delays of GPS signals in their propagation through the atmosphere\cite{garcialorenzo2010}. More recently, encouraging results for the determination of nighttime atmospheric water vapor have been obtained by means of a novel method consisting of measuring the absorption (and modeling the ratios of the absorption), due to atmospheric water vapor, of sunlight reflected from the Moon at four bands centered around 0.95$\mu$m\cite{querel_and_naylor_2011}.

The most recent and, to-date, most successful effort to monitor precipitable water vapor at astronomical sites, utilizing several of the techniques mentioned above, is summarized in Ref. 13, 14 and 28.

Results of long-term statistical medians of PWV for the site of Armazones\footnote{Armazones is located in the Atacama Desert Region, northern Chile.} found in Ref. 8 and reported in Refs. 27 \& 28 differ by about 1 mm (approximately 30\% difference). This difference has been noticed by other authors (see for instance the discussion section in Ref. 29); therefore, it deserves a look at the possible reasons to explain it. The techniques used in these studies involved an assumed water vapor scale height (Ref. 8) and the use of radiosonde soundings launched from the city of Antofagasta\footnote{Antofagasta is a coastal city by the Atacama Desert Region, northern Chile. A radiosonde sounding station, that launches a balloon every day at 12 UT (9 AM local time in summer and 8 AM local time in winter) is situated near to the Antofagasta airport.} (Refs. 27 and 28). The statistical difference between the two studies might be explained by a combination of factors such as: uncertainty in the water vapor scale height (that could potentially introduce a wet-bias in the statistics shown in Ref. 8), a putative dry bias in the radiosonde relative humidity sensors (that could potentially introduce a dry bias in the values reported in Ref. 27), and the spatial and temporal variability of the water vapor field that makes difficult the process of extrapolating the statistics and results of PWV measurements monitored at one site to another a few kilometers away, something potentially affecting the results in Ref. 27. The effects of these important considerations are analyzed in sections 2.1, 2.2 and 2.3, respectively. Overall, these are important considerations to keep in mind when using surface data, spectroscopy and radiometry data as well as radiosonde soundings for the determination of PWV in the atmospheric column.

\section{CONSIDERATIONS WHEN COMPUTING THE PRECIPITABLE WATER VAPOR}

\subsection{Water Vapor Scale Height}

When utilizing astronomical spectrometers and ground-based radiometry, a radiative transfer model is needed together with reasonable assumptions about the state of the atmosphere. In particular, the temperature profile (surface temperature and a temperature profile or lapse rate) and the scale height of the exponential distribution of water vapor are required.  When using surface water vapor density to estimate the integrated water vapor in the atmospheric column, it is assumed that water vapor decreases exponentially with altitude with a given scale height, and also importantly, that the surface layer is weak enough such that the water vapor field is coupled from the free troposphere to the surface. This last assumption is the weakest and the method is found to provide reasonable statistics at sites dominated by high pressure systems, and only when averaging the results using a large number of realizations (see for instance Ref. 8).  As is possible to see, one common requirement of all these methods is knowledge of the water vapor scale height.  Typically, researchers use radiosonde soundings to determine statistically representative values for the mean (or median) of the scale height for a given period or season of interest.

\begin{figure}[t]
   \begin{center}
   \begin{tabular}{c}
   \includegraphics[height=7cm]{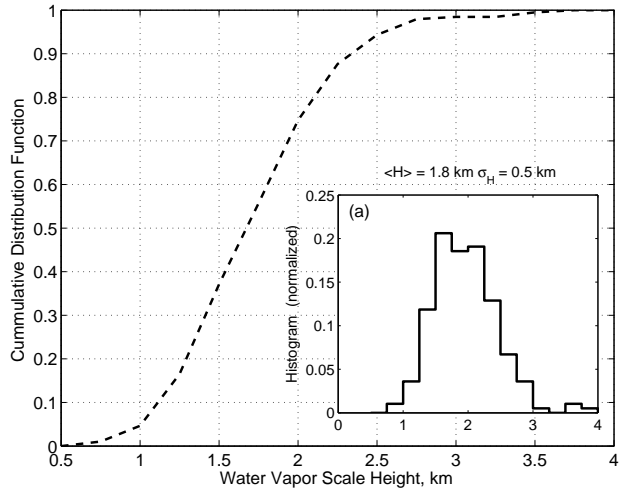}
   \end{tabular}
   \end{center}
   \caption[Water vapor scale height statistics computed from 194 radiosonde soundings launched from Antofagasta in 2005.]
   { \label{fig:figure1}
Water vapor scale height statistics computed from 194 radiosonde soundings launched from Antofagasta in 2005.}
   \end{figure}

Fig.~\ref{fig:figure1} shows the statistics of the water vapor scale height computed from the best exponential fit to the water vapor density (WVD) profile obtained from 194 radiosonde soundings launched from the city of Antofagasta in 2005.  The exponential fit was performed using all data points in the section of the profiles from 2 to 12 km altitude above sea level. It is interesting to note that the histogram (see insert to  Fig.~\ref{fig:figure1}) shows a flat peak in the scale height range from 1.5 to 2.3 km.

PWV can be calculated by integrating the WVD along the vertical axis, see Eq.~\ref{eq:equation1}\footnote{If $\rho_{V}$ and $z'$ are in MKS units then $PWV(z)$ is given in \textit{mm}.} When using radiosonde soundings, the WVD profile, $\rho_{V}(z)$, is obtained from the profiles of temperature, relative humidity and making use of a solution to the Clausius-Clapeyron relationship and the ideal gas equation applied to water vapor (see for instance the first three equations in the expression (3) in Ref. 8). When a true WVD profile is not available then $\rho_{V}$ can be modeled using an exponentially decreasing profile that depends on the value of WVD at a reference altitude ($z_0$, the surface level) and a water vapor scale height ($H$), as show in Eq.~\ref{eq:equation2}. Replacing the exponential profile of WVD (shown in Eq.~\ref{eq:equation2}) in the integral in Eq.~\ref{eq:equation1} and performing the integration between the levels $z_0$ and $z_{max}$ yields an expression for the PWV at the level $z_0$, as shown in Eq.~\ref{eq:equation3}. When $z_{max}$ is a reasonably high altitude above the surface level, then PWV is basically given by the product of the surface WVD (a free parameter in radiative transfer simulations) and $H$, as shown in Eq.~\ref{eq:equation4}. Taking the derivative of Eq.~\ref{eq:equation4} helps one to see that the uncertainties in the surface WVD and $H$ map directly into the uncertainty of the WVD computed from this expression (see Eq.~\ref{eq:equation5}).

\begin{equation}
\label{eq:equation1}
PWV(z) =    \int_z^{z_{max}}\rho_{V}(z')\,dz'
\end{equation}

\begin{equation}
\label{eq:equation2}
\rho_{V}(z) =  \rho_{V_0}\times e^{-\frac{z-z_0}{H}}
\end{equation}

\begin{equation}
\label{eq:equation3}
PWV(z_0) =   \rho_{V}(z_0) \cdot H \cdot \left( 1-e^{-\frac{z_0-z_{max}}{H}} \right)
\end{equation}

\begin{equation}
\label{eq:equation4}
PWV(z_0) =  \rho_{V}(z_0) \cdot H
\end{equation}

\begin{equation}
\label{eq:equation5}
  \frac{dPWV}{PWV} =   \frac{d\rho_V(z_0)}{\rho_V(z_0)} + \frac{dH}{H}
\end{equation}

After replacing the fractional WVD expression in Eq.~\ref{eq:equation5} by its expression in terms of the relative humidity and water vapor pressure (assuming a solution for the Clausius-Clapeyron Equation), it can be shown that the PWV fractional uncertainty can be modeled\footnote{Derivation of the model not shown here.} in terms of the fractional uncertainties of temperature, $T$, relative humidity, $RH$ and water vapor scale height, $H$, as shown in Eq.~\ref{eq:equation6}. It is important to note that in Eq.~\ref{eq:equation6} $T$ is in Kelvins and the term within the circular brackets averages to 19 in the temperature range from 233 K to 313 K (i.e. -40 $^\circ$C to +40 $^\circ$C).This implies that if the temperature readings were to have an uncertainty of 0.5K in an environment of 10 $^\circ$C (283 K), then the temperature uncertainty could contribute to about $19\times 0.5/283 \times 100\% = 3.3\%$ to the uncertainty in PWV.

\begin{equation}
\label{eq:equation6}
\frac{dPWV}{PWV} = ( 3.6 \times 10^{-4}*T^2-0.29*T+71.13) \cdot \frac{dT}{T} + \frac{dRH}{RH} + \frac{dH}{H}
\end{equation}

\begin{figure}[t]
   \begin{center}
   \begin{tabular}{c}
   \includegraphics[height=7cm]{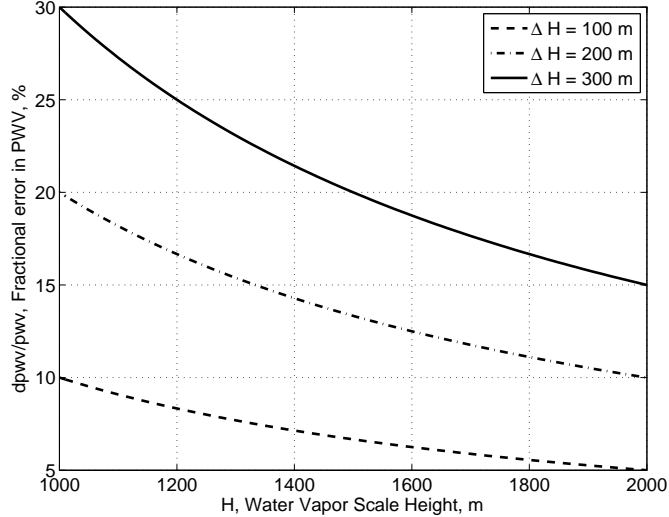}
   \end{tabular}
   \end{center}
   \caption[Absolute fractional error in PWV arising from an absolute error of 100 m (dashed), 200 m (dash-dot) and 300 m (solid), in the water vapor scale height ($H$) as a function of the absolute real value of $H$.]
   { \label{fig:figure2}
   Absolute fractional error in PWV arising from an absolute error of 100 m (dashed), 200 m (dash-dot) and 300 m (solid), in the water vapor scale height ($H$) as a function of the absolute real value of $H$}
   \end{figure}

Assuming no uncertainties in the surface values of temperature and relative humidity, the fractional uncertainty in $H$ maps 1:1 into an uncertainty in PWV. In absolute terms, a constant uncertainty, $\Delta H$, will have a smaller effect as $H$ increases. Fig.~\ref{fig:figure2} shows the effect of a fractional uncertainty in PWV as a function of 100, 200 and 300 m uncertainty in $H$ for different values of scale heights.

The median value of $H$ at Armazones has been found to be on the order of 1.7 km\cite{otarola2010a}. Therefore, fluctuations of $\pm200 m$ in the water vapor scale height introduce an uncertainty of about 12$\%$ in PWV.  However, the insert to Fig.~\ref{fig:figure1} shows a 1-$\sigma$ level in the probability density distribution of $H$ of 0.5 km. Consequently, when using a fixed water vapor scale height of 1.7 km in the derivation of PWV in this region, and taking the actual\footnote{Water vapor in general terms decreases with altitude but does exhibit some fluctuations due to temperature inversions and/or advection processes at different altitudes. The "actual'' scale height (as used here) is only used to refer to the scale height that best fit the decrease of PWV with altitude (at a given time) when assuming an exponential decay model.} scale height to be 1.2 km (-0.5 km from the scale height used in the model) implies an error in the derivation of PWV of about 41$\%$ (model overestimating the PWV). If the actual scale height were 2.2 km (on the other side of the distribution), then the error in PWV reduces to about $22\%$ (this time the model is underestimating). These results show that when deriving PWV using models that require water vapor scale height as an input parameter, all efforts towards determining the most realistic magnitude of $H$ will translate directly into a more accurate determination of PWV.  $H$ changes from day-to-day, and season-to-season, depending on the particular state of the atmosphere, and even changes with time of day.  Possible ways to monitor the actual magnitude of $H$ at a site of interest could include for instance: estimating $H$ from the ratio of WVD measured at two locations with significant differences in altitude\cite{holdaway1996}.  This can also be accomplished using tethersondes equipped with temperature and relative humidity sensors\cite{john2005} at different heights.  Some attempts have been done to infer the water vapor scale height from the rate at which the integrated measurements of water vapor de-correlate with horizontal distance\cite{ruf1997}. It will be advisable to setup a field campaign to test these methods and check the improvement that can be obtained in the estimation of PWV from radiometer data, or from measurements of surface WVD, utilizing a variable $H$ (estimated from additional data) rather than with a fixed value derived from statistical distributions of $H$.

\subsection{Radiosonde: Relative Humidity Dry Bias}

When computing PWV from radiosonde soundings it is absolutely necessary to keep in mind the known problems affecting relative humidity sensors, that in most cases contribute to produce a dry bias in the measurements. Most of the community uses soundings that have been obtained utilizing Vaisala$\copyright$ sensors. Vaisala$\copyright$ uses thin-film capacitive relative humidity sensors with an uncertainty of about 5\% and reproducibility in sounding of about 2\%\cite{vaisalars92}. This thin-film capacitive sensor consists of a polymer layer between the porous electrodes of a capacitor. The sensor changes its capacitance depending on the amount of water vapor molecules that have diffused from the environment into the polymer. The problems reported to affect this type of relative humidity sensors arise from:  calibration model inaccuracy in the response of the sensor with actual environmental temperature, chemical contamination of the polymer (binding sites within the polymer are used by molecules from the sensor's package rendering the sites unavailable for water molecules), sensor aging (sensor response to temperature changes over time affecting its calibration), response time of the sensor changes with temperature (usually resulting in a slower response time with lower temperatures), and also importantly a solar-heating that changes the temperature of the relative humidity sensor with respect to the surrounding environmental temperature and therefore bias the determination of the relative humidity towards drier values. Most of these errors have been identified and studied by researchers interested in improving the accuracy of relative humidity soundings to increase the quality of their measurements for climate studies. These researchers have identified the errors and provided correction algorithms found by simultaneous comparison of the radiosonde humidity soundings to other instruments of known accuracy such as: cryogenic frostpoint hygrometers\cite{milo2004,vomel2007,milo2009}, other calibrated thin-film capacitive polymer sensor\cite{leite2005}, and to microwave radiometers\cite{turner2003}.

Preliminary results from an analysis of the radiosonde soundings launched from the Antofagasta station during the period from July 2005 to November 2010, and using the relative humidity correction algorithm presented in Ref. 35, shows that the median dry bias in the data is on the order of 0.46 mm\footnote{Otarola, presentation at the Astronomical Site Testing Data Conference, held in Valparaiso-Chile on December 1-3, 2010 (http://www.dfa.uv.cl/sitetestingdata/talks.php). Manuscript in preparation.}. Considering no other sources of variability or error, PWV statistics based on the analysis of uncorrected radiosonde soundings launched from the Antofagasta station can potentially be dry biased in a magnitude of about 0.46 mm.

\subsection{Spatial and Temporal Variability of Water Vapor}

Ultimately, when using radiosonde soundings to study the magnitude and fluctuations of precipitable water vapor, it is absolutely necessary to consider the variability of water vapor both in space and in time. In the former case, the farther the sounding station is from the site of interest, the less correlation there will be between the magnitude of the water vapor at the sounding site with that at the site of interest. In the latter case, we have to remember that the atmosphere, specifically in the areas of interest for astronomy, is highly unsaturated with respect to water vapor. Therefore, it is necessary to study how representative of the daily mean (median) conditions is a single measurement of PWV obtained with a radiosonde sounding at a fixed time of day, 0 UT or 12 UT, for instance.

In order to study the variability of the water vapor field as a function of horizontal distance, we made use of a spatial data series of specific humidity obtained by means of instrumented aircraft flying at a constant altitude with a constant bearing. This data comes from the Stratosphere-Troposphere Analyses of Regional Transport (START) experiment\cite{pan2007, bowman2007} obtained with the Gulfstream V (GV) aircraft, and from the TOGA/COARE mission\cite{webster1992, halpern1996} using the Lockheed L-188C ELECTRA (Tail Number N308D) aircraft.

\begin{figure}[b]
   \begin{center}
   \begin{tabular}{c}
   \includegraphics[height=7cm]{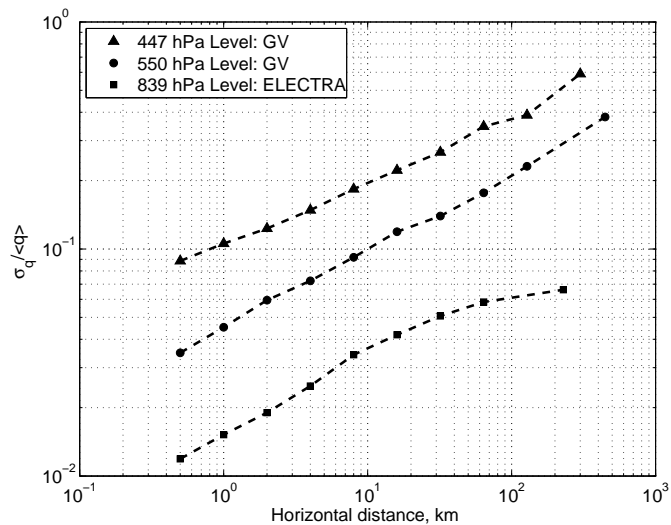}
   \end{tabular}
   \end{center}
   \caption[Fractional variability of specific humidity $\sigma_q/<q>$ as a function of the horizontal path at pressure levels of 447 hPa (triangles), 550 hPa (circles) and 839 hPa (squares). The data was gathered with the GV aircraft for the 447 hPa and 550 hPa, and the Lockheed ELECTRA aircraft for the 839 hPa pressure level.]
   { \label{fig:figure3}
  Fractional variability of specific humidity $\sigma_q/<q>$ as a function of the horizontal path at pressure levels of 447 hPa (triangles), 550 hPa (circles) and 839 hPa (squares). The data was gathered with the GV aircraft for the 447 hPa and 550 hPa, and the Lockheed ELECTRA aircraft for the 839 hPa pressure level.}
   \end{figure}

Fig.~\ref{fig:figure3} shows the ratio of the standard deviation, $\sigma_q$, to the mean, $<q>$, of specific humidity computed for variable horizontal distances ranging from 500 m to more than 200 km. The spatial resolution of the specific humidity measurements is 220 m for the GV aircraft data and 100 m for the Lockheed ELECTRA aircraft data. Consequently, in a 10 km path there are about 45 and 100 values of $q$ in the data series from the GV and the Lockheed ELECTRA aircrafts, respectively.

There is much to learn from Fig.~\ref{fig:figure3}, however, here the emphasis is on the spatial variability of the humidity field. For example, the 550 hPa level data shows that in a 100 km horizontal path (comparable to the distance from the Antofagasta radiosonde sounding station to the Armazones site), the 1-$\sigma$ level of $q$ is about 23$\%$ of its mean value. In other words, any two measurements of specific humidity separated by 100 km at an altitude corresponding to the 550 hPa pressure level could differ by 23$\%$ with a 66.7\% probability level. If the two measurements were separated by a distance of 10 km, the 1-$\sigma$ variability decreases to 10$\%$.  The message here is that, the statistics of PWV obtained from the analysis of well-calibrated radiosonde soundings launched from the city of Antofagasta might differ in the order of 10\%-30\% at a location such as Armazones located 100 km away from Antofagasta, this only accounting for the horizontal variability of the water vapor field.

Importantly, Fig.~\ref{fig:figure3} also shows that the variability of the humidity field along a given horizontal path depends on the altitude within the troposphere. For a given altitude, there might also be different levels of variability with geographic region and season. Fig.~\ref{fig:figure3} was computed using data from winter-mid-latitude and summer-equatorial regional conditions for the case of the GV and Lockheed aircraft data, respectively. Fig.~\ref{fig:figure3} shows how dramatic the spatial differences can be across the humidity field, reminding us to exercise caution when employing extrapolated results from radiosonde soundings located many kilometers away from the site being studied.

\begin{figure}[t]
   \begin{center}
   \begin{tabular}{c}
   \includegraphics[height=7cm]{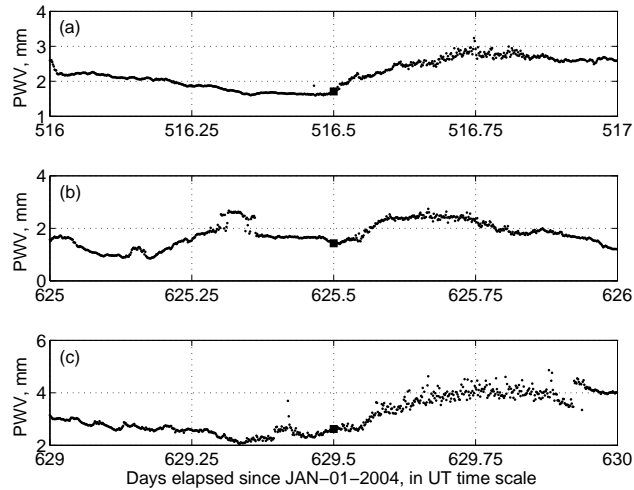}
   \end{tabular}
   \end{center}
   \caption[Time series of PWV at Armazones: (a) day 516, (b) day 625, and (c) day 629 elapsed since Jan-01-2004. The square at noon UT is to emphasize the single measurement that could be obtained from a 12 UT radiosonde sounding.]
   { \label{fig:figure4}
  Time series of PWV at Armazones: (a) day 516, (b) day 625, and (c) day 629 elapsed since Jan-01-2004. The square at noon UT is to emphasize the single measurement that could be obtained from an in-situ 12 UT radiosonde sounding.}
   \end{figure}

Another important consideration is how representative the daily statistics can be of a one-time PWV measurement at a given site. Fig.~\ref{fig:figure4} shows the PWV time series for three different days for the site of Armazones. The PWV was derived from surface measurements of temperature and relative humidity properly converted into surface WVD as shown in Ref. 8. Table ~\ref{tab:table1}, shows the statistical values of PWV for the three days in Fig.~\ref{fig:figure4}. The statistics show that in the first case, see Fig.~\ref{fig:figure4}(a), a single measurement at 12 UT (like one which could have been obtained from an in-situ radiosonde sounding) matches better the minimum value of the 24-h time series. While in the second and third cases, see Fig.~\ref{fig:figure4}(b \& c), the 12 UT measurement match within 10\% the nightime median of PWV, they show a varying degree of accuracy when compared to the overall daily medians. Consequently, it is important to look at the time variability of precipitable water vapor before assigning a single measurement at a fixed time to represent the day-time, night-time or overall daily conditions of a site.

\begin{table}[h]
\caption{PWV statistics for the time series shown in Fig.~\ref{fig:figure4} }
\label{tab:table1}
\begin{center}
\begin{tabular}{c | c | c | c | c | c | c} 
\hline
\rule[-1ex]{0pt}{3.5ex} Day No & min & max & night & day & daily & 12 UT \\
\rule[-1ex]{0pt}{3.5ex}            & pwv & pwv & median & median & median & pwv \\
\rule[-1ex]{0pt}{3.5ex}            & [mm] & [mm] & [mm] & [mm] & [mm] & [mm] \\
\hline
\rule[-1ex]{0pt}{3.5ex} 516 & 1.67 & 3.24 & 1.95 & 2.60 & 2.20 & 1.71 \\
\hline
\rule[-1ex]{0pt}{3.5ex} 625 & 0.90 & 2.74 & 1.60 & 2.00 & 1.70 & 1.43 \\
\hline
\rule[-1ex]{0pt}{3.5ex} 629 & 2.10 & 4.86 & 2.60 & 3.81 & 2.90 & 2.62 \\
\hline
\end{tabular}
\end{center}
\end{table}

\section{CONCLUSIONS}

In conclusion, the determination of the integrated water vapor in the atmospheric column from either the radiometry technique (in some cases) or from surface measurements of water vapor density do assume a model on which water vapor decays exponentially with altitude with a given scale height. If all other parameters are known to a high accuracy, then the fractional uncertainty in the magnitude of the water vapor scale height will translate directly, in a 1:1 relationship, to the uncertainty in the magnitude of PWV. As shown in Section 2.1 (Fig.~\ref{fig:figure2}) a 100 m uncertainty in the value of the scale height used in the models has a more significant effect when the actual scale height is smaller.  The effect is that, in the ideal case of water vapor distribution with altitude, if the actual value of the water vapor scale height is 1.2 km and the model uses 1.5 km, then the model is overestimating the magnitude of PWV for about (1.5-1.2)/1.2 $\sim$ 25\%. On the other hand, if actual value of the scale height is 1.8 km (300 meters uncertainty as in the first case), the models will be underestimating PWV by about (1.5-1.8)/1.8 $\sim$ 17\%. Consequently, using a constant value for the water vapor scale height in radiative transfer models, or in estimating water vapor from surface measurements, effectively introduces a wet- or a dry bias from day-to-day or season-to-season depending on whether the actual scale height is smaller or larger than that used in the calculations. Considering that the spread in the water vapor scale height is not negligible, as shown in the insert to Fig.~\ref{fig:figure1}, the dry and wet biases can easily be on the order of 20\% to 40\% in magnitude. When calculating the global statistics (overall mean value for instance), however, the effect of wet and dry bias decreases, but might not cancel completely. This is due to an asymmetry in the ratio of the difference between the $H$ used in the model and its actual value (above or below that used in the model) at a given time. Monte Carlo simulations, where a scale height for a given day is picked randomly from a water vapor scale height distribution of known mean and standard deviation, and for each WVD data point available in the Armazones data series, shows that the overall median PWV of 3.2 mm for Armazones\cite{otarola2010a} might have been overestimated by about 2.1\% (i.e. by  about 0.01 mm). It is important to keep in mind that the actual atmosphere shows temperature inversions that tend to trap water vapor below the ceiling of the inversion preventing mixing of water vapor with the free troposphere. Depending on the strength of the temperature inversion layers, this introduces a departure from the assumed exponential-decay model and becomes a large source of error. Exploring and testing methods to determine the water vapor scale height and use it in the radiative transfer calculations will increase our understanding of the PWV variations observed on both short and long time-scales, from day-to-day and from season-to-season.

In the case of using spectroscopy or radiometry data, an alternative to using an exponential model for the water vapor distribution in the radiative transfer model could be that of using normalized profiles of temperature and PWV that have been obtained, for the site of interest, from the analysis of radiosonde soundings.   The normalized temperature profile\footnote{The temperature profile is normalized to its surface value.} will include the average location and strength of temperature inversion layers, and can be scaled to the actual conditions using a measurement of the surface temperature. Similarly, the normalized PWV\footnote{The water vapor density profile is normalized to its integral value.} profile will be, this way, a more realistic representation of the distribution of water vapor at the site of interest including the effects of layering that might be the result of atmospheric circulation or temperature inversions. In this approach the radiative transfer model can solve for the PWV (by scaling the whole water vapor density profile) until the modeled brightness temperature (astronomical fluxes of standard stars) matches that observed by the radiometers (spectrographs).

On the other hand, Sections 2.2 and 2.3 shows that when using water vapor statistics derived from radiosonde soundings, it is absolutely necessary to keep in mind that the atmosphere is highly unsaturated with respect to water vapor, and that there is plenty of room for spatial and temporal fluctuations. Also, importantly and as reported in the technical literature, relative humidity sensors are known to be affected by physical and environmental factors that, in the overall, tend to introduce a dry bias in the humidity data series.

A combination of the effects introduced in this work might help to explain the differences found between the statistics of precipitable water vapor at  Armazones reported by the TMT (Ref. 8) and the E-ELT (Ref. 27 \& 28) site testing groups, of 3.2 mm and 2.1 mm, respectively. It is important to clarify that though the analysis of PWV for Armazones, based on surface WVD, conducted by the TMT group gave an overall (day \& night) value of 3.2 mm, the median of night-time conditions was found to be 2.9 mm. Therefore, the value adopted for site comparisons purposes was 2.9 mm which is also the result from the analysis of GOES satellite data.

\acknowledgments     

The author thanks Dr. Laura Pan (UCAR) for making available the GV aircraft data gathered during the START experiment, and to Dr. Ron Ruth (RAF) and Dr. Yolande Serra (University of Arizona) for helping in getting access to the ELECTRA aircraft data gathered during the TOGA-COARE mission.
A. Ot{\'a}rola gratefully acknowledge the support of the TMT partner institutions. They are the Association of Canadian Universities for Research in Astronomy (ACURA), the California Institute of Technology and the University Of California. This work was supported as well by the Gordon and Betty Moore Foundation, the Canada Foundation for Innovation, The National Research Council of Canada, the Natural Sciences and Engineering Research Council of Canada, The British Columbia Knowledge Development Fund, the Association of Universities for Research in Astronomy (ACURA) and the U.S. National Science Foundation.


\end{document}